\documentclass{article}

\usepackage{arxiv}

\usepackage[utf8]{inputenc} 
\usepackage[T1]{fontenc}    
\usepackage[hidelinks]{hyperref}       
\usepackage{url}            
\usepackage{booktabs}       
\usepackage{amsfonts}       
\usepackage{nicefrac}       
\usepackage{microtype}      
\usepackage{lipsum}		
\usepackage{graphicx}
\usepackage[numbers]{natbib}
\usepackage{doi}
\usepackage{multirow, longtable}
\usepackage{float}

\title{Multiple type I error concepts for clinical trials with overlapping populations}

\author{{Remi Luschei} \\
	Competence Center for Clinical Trials Bremen \\
	Institute for Statistics\\
	University of Bremen\\
	\texttt{rluschei@uni-bremen.de} \\
	\And
	{Werner Brannath} \\
	Competence Center for Clinical Trials Bremen\\
	Institute for Statistics \\
	University of Bremen\\
	\texttt{brannath@uni-bremen.de} \\
}

\renewcommand{\headeright}{}
\renewcommand{\undertitle}{}
\hypersetup{
pdftitle={Multiple type I error concepts for clinical trials with overlapping populations},
pdfauthor={Remi Luschei, Werner Brannath},
}

\usepackage{amsmath, amsfonts, mathtools, csquotes, url, bbm, tikz, bm, amsthm, subcaption}
\usepackage[titletoc,toc,page]{appendix}
\newcommand{\Pop}{\mathcal{P}}
\renewcommand{\P}{\mathbb{P}}
\newcommand{\PWER}{\text{PWER}}
\newcommand{\FWER}{\text{FWER}}
\newcommand{\PWERP}{\text{PWER-P}}
\newcommand{\PWERU}{\text{PWER-U}}
\newtheorem{theorem}{Theorem}
\theoremstyle{definition}
\newtheorem{example}{Example}

\usepackage{setspace}
\begin{document}
\maketitle

\begin{abstract}
The population-wise error rate (PWER) was introduced as a more liberal alternative to the family-wise error rate (FWER) for clinical trials with multiple, overlapping patient populations. These trials are particularly relevant in personalized medicine, which aims to find therapies tailored to specific patient subgroups. By controlling an average multiple type I error probability over all population strata, the PWER can substantially improve statistical power. However, one disadvantage of this concept is that the error probability for a given population can strongly depend on the presence or absence of other populations included in the analysis. To address this issue, we propose two modifications of the PWER that enforce individual error control either for all target populations, or for all possible unions of target populations. We call these approaches the PWER over the populations (PWER-P) and the PWER over population unions (PWER-U). We investigate the properties of these new error rates and compare them with the PWER and FWER in terms of type I error control and power.
\end{abstract}


\section{Introduction} \label{sec:intro}

The increasing complexity of modern clinical trials has brought renewed attention to the question of multiplicity control. Advances in personalized medicine, the growing focus on rare diseases, and innovative trial designs have led to settings where treatment effects are evaluated simultaneously across several patient populations. The aim is to develop therapies that are tailored to specific patient subgroups which are identified by certain genetic or clinical biomarkers. Notable examples include umbrella and basket trials in oncology, as well as adaptive platform and enrichment trials (see e.g.\ \cite{park2020, stallard2019, baldi2023}).

In confirmatory studies, control of the family-wise error rate (FWER) remains the most widely accepted approach to limit multiple type I errors. It bounds the probability of making at least one false positive conclusion across all tested hypotheses, and is a central element of current regulatory practice. However, in studies with multiple populations, the FWER may not always reflect the number of decisions that are actually relevant to all patients. Instead, it represents a \emph{maximal} risk for future patients of being exposed to inefficient treatments, which can make it overly restrictive. To illustrate this, consider a study with two overlapping populations, $\Pop_1$ and $\Pop_2$, where two targeted treatments, $T_1$ and $T_2$, are tested for efficacy, respectively. This corresponds to testing the null hypotheses $H_i\colon T_i \text{ is not effective in } \Pop_i$, for $i=1,2$. The patients in the population overlap, $\Pop_{\{1,2\}} = \Pop_1 \cap \Pop_2$, could  be exposed to future risks both in case of an incorrect rejection of $H_1$ and of $H_2$. However, the patients in the complements, $\Pop_{\{i\}} = \Pop_i \setminus \Pop_{\{1,2\}}$, $i = 1,2$, are only affected by the hypothesis $H_i$, and are therefore exposed to lower risks. For this reason, controlling the FWER may unnecessarily sacrifice power in this situation.

As an alternative to the FWER, \citet{brannath2023} introduced the population-wise error rate (PWER), which is a more liberal type I error rate and corresponds to the \emph{average} risk for future patients of being exposed to inefficient treatments. In the setting with two populations, it is defined as 
\begin{align} \label{eq:pwer-2pops}
\PWER = \pi_{\{1\}} \P(\text{falsely reject $H_1$}) + \pi_{\{2\}} \P(\text{falsely reject $H_2$})+ \pi_{\{1,2\}} \P(\text{falsely reject $H_1$ or $H_2$}),
\end{align}
where $\pi_J$, for every nonempty $J \subseteq \{1,2\}$, denotes the prevalence of $\Pop_J$ within the overall population $\Pop_1 \cup \Pop_2$. In general, with an arbitrary number of target populations $\Pop_i$, $i \in I = \{1, \dots, m\}$, and corresponding null hypotheses $H_i$, the PWER is defined as 
\begin{align} \label{eq:pwer-def}
\PWER = \sum_{J \subseteq I} \pi_J \P(\text{falsely reject any $H_j$ for $j \in J$}), 
\end{align}
where $\pi_J$ denotes the prevalence of the stratum $\Pop_J = \cap_{j \in J} \Pop_j \setminus \cup_{k \not\in J} \Pop_k$. If we use univariate test statistics $T_i$ for $H_i$ and reject $H_i$ if $T_i \geq c_i$ for some critical value $c_i$, the PWER is usually maximized under the global null hypothesis, that is, when all $H_i$ are true \cite{luschei2025prevalence}. Therefore, when the joint distribution of the test statistics under the global null hypothesis is known, the PWER can be controlled at a significance level $\alpha \in (0,1)$ by choosing the critical values $c_1,\ldots, c_m$ such that $\PWER = \sum_{J \subseteq I} \pi_J(1-F_J(\boldsymbol{c}_J)) \leq \alpha$, where $F_J$ denotes the joint distribution function of $(T_j)_{j \in J}$ evaluated in $\boldsymbol{c}_J = (c_j)_{j \in J}$. Compared with FWER control, controlling the PWER provides the advantage of enabling higher power and lower sample sizes, as has been demonstrated in \cite{brannath2023}.

However, using the PWER may also lead to undesirable properties in certain situations. Consider, for example, a study with two disjoint populations of equal prevalence, i.e.\ $\pi_{\{1\}} = \pi_{\{2\}} = 1/2$ and $\pi_{\{1,2\}} = 0$. In this case, the PWER under the global null hypothesis becomes $\PWER = \P(\text{reject $H_1$})/2 + \P(\text{reject $H_2$})/2$. Consequently, testing both hypotheses at level $\alpha$ would ensure that $\PWER \leq \alpha$. However, one could also test $H_1$ at level $3\alpha/2$ and $H_2$ at level $\alpha/2$, which still yields $\PWER \leq \alpha$, but disadvantages population $\Pop_1$ in terms of type I error control. This shows that PWER control alone does not uniquely determine how the type I error level is distributed across the populations, and that it may allow for highly imbalanced allocations. To address this issue, \citet{brannath2023} recommend testing all hypotheses at identical local levels, which avoids arbitrary allocations of error levels across populations.

\begin{figure}[t]
\centering
\begin{minipage}{0.48\textwidth}
\centering
\begin{tikzpicture}
  \draw (0cm,0cm) ellipse[x radius=1.5cm,y radius=1.3cm] node at (-0.5,0) {$\Pop_{\{1\}}$};
  \draw (1.75cm,0cm) ellipse[x radius=1.5cm,y radius=1.3cm] node at (2.25,0) {$\Pop_{\{2\}}$};
  \node at (0.85,0) {$\Pop_{\{1,2\}}$};
\node at (-0.5, -1.8) {$\Pop_1$};
\node at (2.25, -1.8) {$\Pop_2$};
\end{tikzpicture}
\caption{Two overlapping populations}
\label{fig:2opops}
\end{minipage}
\begin{minipage}{0.48\textwidth}
\centering
\begin{tikzpicture}
\draw (0cm,0cm) ellipse[x radius=1.5cm,y radius=1.3cm]
      node at (-0,0) {$\Pop_{\{1\}}$};
\draw (3.5cm,0cm) ellipse[x radius=1.5cm,y radius=1.3cm]
      node at (3.5,0) {$\Pop_{\{2\}}$};
\node at (0cm,-1.8cm) {$\Pop_1$};
\node at (3.5cm,-1.8cm) {$\Pop_2$};
\end{tikzpicture}
\caption{Two disjoint populations}
\label{fig:2dpops}
\end{minipage}
\end{figure}

A further aspect to consider with PWER control is that the selection of target populations can affect the error probability of a given population. To illustrate this, we again consider a study with two overlapping populations, where the same experimental treatment is compared to a control treatment. We further assume normally distributed data, with strata-specific means, but homogeneous variances. As shown in \cite{brannath2023}, under these conditions the hypotheses $H_i$ can be tested with test statistics that jointly follow a bivariate normal distribution with correlation $\rho = \pi_{\{1,2\}}/[(\pi_{\{1\}}+\pi_{\{1,2\}})(\pi_{\{2\}}+\pi_{\{1,2\}})]^{1/2}$. Thus, when using a common critical value $c$ for both tests, the PWER can be controlled by solving the equation 
\begin{align}\label{eq:iex2}
\PWER = (\pi_{\{1\}} + \pi_{\{2\}})(1-\Phi(c)) + \pi_{\{1,2\}}(1-\Phi_\rho(c,c)) = \alpha
\end{align}
for $c$, where $\Phi$ is the standard normal distribution function and $\Phi_\rho$ is the bivariate normal distribution function with correlation $\rho$. If we assume that all strata have the same size, i.e.\ $\pi_{\{1\}} = \pi_{\{2\}} = \pi_{\{1,2\}} = 1/3$, then we get $\rho = 0.5$. For $\alpha = 0.025$, solving equation \eqref{eq:iex2} yields $c = 2.06$. Now, suppose that a third population $\Pop_3$, disjoint from $\Pop_1$ and $\Pop_2$, is included in the study. In this case, the PWER becomes 
\begin{align*}
\PWER = (\pi_{\{1\}} + \pi_{\{2\}} + \pi_{\{3\}})(1-\Phi(c)) + \pi_{\{1,2\}}(1-\Phi_\rho(c,c)).
\end{align*}
If we assume that $\Pop_3$ has a prevalence of $0.75$, and that the other strata remain equally sized, such that $\pi_{\{1\}}=\pi_{\{2\}}=\pi_{\{1,2\}}=1/12$, the correlation between $\Pop_1$ and $\Pop_2$ remains $\rho = 0.5$. Controlling the PWER at the level $\alpha = 0.025$ then gives $c = 1.99$. Thus, although the testing procedure and the PWER level remain unchanged, the population-wise error probabilities change. For $\Pop_1$, the probability that a randomly selected patient is affected by at least one type I error equals $(1-\Phi(c))/2 + (1-\Phi_{0.5}(c,c))/2$, which is 0.028 when $c = 2.06$ and 0.033 when $c = 1.99$. Of course, the same also applies to $\Pop_2$. Hence, under PWER control, the population-wise error probability of a given target population depends on the overall study configuration and may even exceed the nominal significance level.

More broadly, these observations highlight a certain risk of misuse associated with the PWER: certain populations may be disadvantaged in terms of type I error control, either through an unequal allocation of error levels, or through the inclusion or exclusion of other target populations. To address these issues, in this paper we propose to modify the PWER by requiring individual error control for all target populations in which hypotheses are being tested. We call this control of the PWER over the populations (PWER-P). As an alternative, we also propose to control the PWER across all possible unions of target populations (PWER-U). 

We will investigate the properties of these new error rates and compare them to the PWER and FWER in terms of type I error control and power. Moreover, it will be shown that the practical handling of the PWER-P and PWER-U is very similar to that of the PWER and FWER. For example, they can be controlled under the same conditions, and corresponding confidence intervals can similarly be constructed. All simulations performed are implemented in R. The script files are available at the link: \url{https://github.com/rluschei/pwer-p}

\section{The population-wise error rate over populations and population unions} \label{sec:def}

In this section, we formally introduce the PWER-P and PWER-U, we show how to control them at a given significance level, and illustrate the control for different study settings.

\subsection{Definitions}

The PWER extends naturally to arbitrary unions of the disjoint population strata $\Pop_J$. To this end, let $C \subseteq \mathcal{C} \coloneqq 2^I \setminus \{\emptyset\}$ be a collection of strata indices, define $\Pop^C = \cup_{J \in C} \Pop_J$ and denote by $\pi^C = \sum_{J \in C} \pi_J$ the prevalence of $\Pop^C$. If $\pi^C > 0$, the PWER for $\Pop^C$ can be defined as 
\begin{align} \label{eq:pwerc}
\PWER_C = \sum_{J \in C} \frac{\pi_J}{\pi^C} \FWER_J,
\end{align}
where $\FWER_J = \P(\text{falsely reject any $H_j$ for $j \in J$})$ denotes the strata-wise error rate of $\Pop_J$. In particular, for $C = \mathcal{C}$, this definition coincides with the overall PWER defined in (\ref{eq:pwer-def}). 

The PWER of an individual target population $\Pop_i$, $i \in I$, can be obtained from (\ref{eq:pwerc}) by considering the collection $C_i = \{J \subseteq I: i \in J\}$ containing all strata belonging to this population. We define the \emph{population-wise error rate over the populations (PWER-P)} as the largest PWER reached across all target populations, \[\PWERP = \max_{i \in I^+} \PWER_{C_i}, \quad I^+ = \{i \in I: \pi^{C_i}>0\}.\]

Analogously, for a given union of target populations $\cup_{u \in U} \Pop_u$, $U \in \mathcal{C}$, we consider the collection $C_U = \{J \subseteq I: U \cap J \neq \emptyset\}$. This leads to the \emph{population-wise error rate over population unions (PWER-U)}, which is defined by \[\PWERU = \max_{U \in \mathcal{U}^+ }\PWER_{C_U}, \quad \mathcal{U}^+ = \{U \in \mathcal{C}: \pi^{C_U}>0\}.\]

\begin{example} \label{ex:2pops}
We consider a study with two overlapping populations $\Pop_1$ and $\Pop_2$, as already regarded in the introduction of this paper. In this case, the PWERs of the populations $\Pop_1$ and $\Pop_2$ are given by
\begin{align*}
\PWER_{C_1} &= \frac{\pi_{\{1\}}\P(\text{falsely reject $H_1$})+\pi_{\{1,2\}}\P(\text{falsely reject $H_1$ or $H_2$})}{\pi_{\{1\}} + \pi_{\{1,2\}}}, \\
\PWER_{C_2} &= \frac{\pi_{\{2\}}\P(\text{falsely reject $H_2$})+\pi_{\{1,2\}}\P(\text{falsely reject $H_1$ or $H_2$})}{\pi_{\{2\}} + \pi_{\{1,2\}}},
\end{align*}
and $\PWERP = \max\{\PWER_{C_1}, \PWER_{C_2}\}$. The PWER-U additionally includes the overall PWER from Equation~(\ref{eq:pwer-2pops}) in the maximization. But since $\PWER \leq \max\{\PWER_{C_1}, \PWER_{C_2}\}$, it follows that $\PWERP = \PWERU$ in this example.
\end{example}

The following theorem addresses the relations between the different error rates.

\begin{theorem} \label{thm1}
It holds $\text{PWER-P} \leq \text{PWER-U} \leq \FWER$. If, in addition, the condition 
\begin{align} \label{cond1}
\forall J, J' \subseteq I: \; |J| < |J'|, \; \pi_J \neq 0, \; \pi_{J'} \neq 0 \;\implies \; \FWER_J \leq \FWER_{J'}
\end{align} 
is satisfied, then $\PWER \leq \text{PWER-P}$ also holds.
\end{theorem}
The proof of Theorem~\ref{thm1} can be found in Appendix~\ref{app1}. We note that the condition~(\ref{cond1}) is always satisfied with two populations (Example~\ref{ex:2pops}).

\subsection{Control of the PWER-P and PWER-U} \label{sec:control}

The PWER-P and PWER-U can be controlled similarly to the PWER. For each $i \in I$, let $T_i$ be a test statistic for testing $H_i$ against a right-sided alternative, and assume that the joint distribution of $(T_i)_{i \in I}$ is known under the global null hypothesis. The PWER-P or the PWER-U can then be controlled at the level $\alpha \in (0,1)$ by choosing a vector of critical values $\boldsymbol{c} = (c_i)_{i \in I}$ to ensure
\begin{align} 
\PWERP(\boldsymbol{c}) &= \max_{i \in I^+} \PWER_{C_i}(\boldsymbol{c}) = \max_{i \in I^+} \sum_{J \in C_i} \frac{\pi_J}{\pi^{C_i}}\left(1-F_J(\boldsymbol{c}_J)\right) = \alpha, \label{eq:pwerp-crit}\\
\text{or} \quad \PWERU(\boldsymbol{c}) &= \max_{U \in \mathcal{U}^+} \PWER_{C_U}(\boldsymbol{c}) = \max_{U \in \mathcal{U}^+} \sum_{J \in C_U} \frac{\pi_J}{\pi^{C_U}}\left(1-F_J(\boldsymbol{c}_J)\right) = \alpha, \label{eq:pweru-crit}
\end{align}
respectively. As above, $F_J$ denotes the joint distribution function of $(T_j)_{j\in J}$ under the global null hypothesis, evaluated in $\boldsymbol{c}_J = (c_j)_{j \in J}$. Similarly, critical values for left-sided tests can also be calculated by replacing $1-F_J(\boldsymbol{c}_J)$ with $F_J(\boldsymbol{c}_J)$ in the equations above. In the following, we restrict our attention to right-sided tests for simplicity.

In general, it is not possible to exhaust all population-wise errors $\PWER_{C_i}(\boldsymbol{c})$ or all union-wise errors $\PWER_{C_U}(\boldsymbol{c})$ simultaneously at the level $\alpha$. The reason is that the critial values $c_i$ depend on the populations $\Pop_i$ and may appear in multiple of the strata-wise error rates $1-F_J(\boldsymbol{c}_J)$, $J \subseteq I$, which is the case when at least two populations overlap. An example will be given in Section~\ref{sec:2pop}.

Alternatively, PWER-P and PWER-U control can also be expressed in terms of adjusted p-values. The adjusted p-value for the hypothesis $H_i$ is defined as the smallest significance level $\alpha$ at which $H_i$ is rejected under the corresponding PWER-P or PWER-U controlling test. When a common critical value $c$ is used for all hypotheses, it can be obtained by evaluating expression~(\ref{eq:pwerp-crit}) or~(\ref{eq:pweru-crit}), respectively, at $c=t_i$. 

Possible choices of test statistics $(T_i)_{i \in I}$ for different settings with normally distributed observations and homogeneous or heterogeneous strata-wise variances are presented in \cite{luschei2025prevalence}, Section~3. 

In the following subsections, we illustrate the error control in different examples.

\subsection{Two overlapping populations} \label{sec:2pop}

For two populations (see Example~\ref{ex:2pops}), with the notation from Section~\ref{sec:control} we have
\begin{align*}
\PWER_{C_1}(\boldsymbol{c}) &= \frac{\pi_{\{1\}}(1-F_{\{1\}}(c_1))+\pi_{\{1,2\}}(1-F_{\{1,2\}}(c_1, c_2))}{\pi_{\{1\}} + \pi_{\{1,2\}}}, \\
\PWER_{C_2}(\boldsymbol{c}) &= \frac{\pi_{\{2\}}(1-F_{\{2\}}(c_2))+\pi_{\{1,2\}}(1-F_{\{1,2\}}(c_1, c_2))}{\pi_{\{2\}} + \pi_{\{1,2\}}}.
\end{align*}
We can directly see that exhausting $\PWER_{C_1}(\boldsymbol{c})$ and $\PWER_{C_2}(\boldsymbol{c})$ simultaneously at level $\alpha$ is, in general, impossible as they both depend on $c_1$ and $c_2$. When $T_1$ and $T_2$ have the same marginal distribution, one could e.g.\ use a common critical value $c = c_1 = c_2$ and control the PWER-P by solving the equation $\PWERP(c) = \alpha$ for $c$.

\subsection{Disjoint populations} \label{sec:disj_pop}

When all target populations $\Pop_1, \dots, \Pop_m$ are pairwise disjoint, i.e.\ $\Pop_i \cap \Pop_j = \emptyset$ for $i \neq j$, we find that \[\PWERP(\boldsymbol{c}) = \PWERU(\boldsymbol{c}) = \max_{i\in I^+} \left(1-F_{\{i\}}(c_i)\right).\] Thus, both the PWER-P and PWER-U can be controlled at level $\alpha$ by determining the critical values from the conditions $ \PWER_{C_i}(\boldsymbol{c}) = 1-F_{\{i\}}(c_i) = \alpha$, for all $i \in I^+$. In particular, it makes sense to use heterogeneous critical values in this case, because all population-wise and union-wise error rates can fully be exhausted. The $\PWER = \sum_{i\in I} \pi_{\{i\}}(1-F_{\{i\}}(c_i))$ is then also controlled at level $\alpha$. However, unlike the PWER-P and PWER-U, the latter also allows for unequal allocations of error levels across populations, which may not be desirable (see Section~\ref{sec:intro}).

When only certain populations are disjoint from others, exhausting error control can be applied at least to those populations. For example, consider again a study with two overlapping populations and a third population that is pairwise disjoint from both. In this setting, the PWER-P and PWER-U can be controlled by computing $c_1 = c_2$ as described in Section~\ref{sec:2pop}, while $c_3$ is obtained by solving $1-F_{\{3\}}(c_3) = \alpha$.

\subsection{Nested populations} \label{sec:nested-pops}

For nested populations $\Pop_1 \supseteq \Pop_2 \supseteq \dots \supseteq \Pop_m$ (see Figure~\ref{fig:nest}), which occur for example in enrichment trials, both PWER-P and PWER-U coincide with the FWER. The reason is that the stratum $\Pop_{\{1, \dots, m\}} = \cap_{i=1}^m \Pop_i$, whose patients are affected by all test decisions, is then identical to the target population $\Pop_m$. Consequently, it has the largest strata-wise FWER and the largest average multiple type I error probability among all populations and population unions. Therefore, in this situation one may prefer to still use the overall PWER, which takes the form \[\PWER(\boldsymbol{c}) = \sum_{i=1}^m \pi_{\{1, \dots, i\}} (1-F_{\{1, \dots, i\}}(c_1, \dots, c_i)).\]

\begin{figure}[t]
\begin{minipage}{0.48\textwidth}
\centering
\begin{tikzpicture}
  \draw (0cm,0cm) ellipse[x radius=3.2cm,y radius=1.3cm] 
        node at (-2.5,0) {$\Pop_{\{1\}}$};
  \draw (.4cm,0cm) ellipse[x radius=2.4cm,y radius=1.1cm] 
        node at (-1.2,0) {$\Pop_{\{1,2\}}$};
  \draw (.8cm,0cm) ellipse[x radius=1.4cm,y radius=0.8cm] 
        node at (0.9,0) {$\Pop_{\{1,2,3\}}$};
  \node at (-2.5,-1.6) {$\Pop_1$};
  \node at (-1.2,-1.6) {$\Pop_2$};
  \node at (0.9,-1.6) {$\Pop_3$};
\end{tikzpicture}
\caption{Three nested populations}
\label{fig:nest}
\end{minipage}
\begin{minipage}{0.48\textwidth}
\centering
\begin{tikzpicture}
\draw (0,0) ellipse(1.8cm and 1.cm);
\draw (-1,0.9) ellipse(1.8cm and 1.cm);
\draw (1,0.9) ellipse(1.8cm and 1.cm);
\node at (-2.8,0.1) {$\Pop_1$};
\node at (2.8, 0.1) {$\Pop_2$};
\node at (-1.75,1.1) {$\Pop_{\{1\}}$};
\node at (1.75,1.1) {$\Pop_{\{2\}}$};
\node at (0,-0.4) {$\Pop_{\{3\}}$};
\node at (0,1.275) {$\Pop_{\{1,2\}}$};
\node at (0,0.625) {$\Pop_{\{1,2,3\}}$};
\node at (0, -1.4) {$\Pop_3$};
\node at (-1,0.25) {$\Pop_{\{1,3\}}$};
\node at (1.15,0.25) {$\Pop_{\{2,3\}}$};
\end{tikzpicture} 
\caption{Three overlapping populations}
\label{fig:3pops}
\end{minipage}
\end{figure}

\subsection{An example where $\PWERP \neq \PWERU$}

In the examples presented so far, the PWER-P always coincides with the PWER-U. However, this equality does not generally hold. To illustrate this, we consider a study with three overlapping populations (see Figure~\ref{fig:3pops}). Let the prevalences be given by $\pi_{\{1\}} = \pi_{\{3\}} = \pi_{\{1,2,3\}} = 0.01$, $\pi_{\{1,2\}} = \pi_{\{1,3\}} = 0.22$, $\pi_{\{2\}} = 0.42$, $\pi_{\{2,3\}} = 0.11$. We further assume that the test statistics follow a multivariate normal distribution, with correlations given by 
\begin{align} \label{eq:corr}
\text{Cov}(T_i, T_j) = \frac{\sum_{J \subseteq I: \, i,j \in J} \pi_J}{\sqrt{\left(\sum_{J \subseteq I: \, i \in J} \pi_J\right)\left(\sum_{J \subseteq I: \, j \in J}\pi_J\right)}}.
\end{align}
This is the case, at least asymptotically, when the same treatment is investigated in all populations, and when the observations are assumed to have homogeneous variances across all strata \cite{luschei2025prevalence}. For the critical values $c_1= 2$, $c_2 = 1.8$, $c_3 = 2$, we then obtain $\PWERP =  0.0476 < \PWERU = 0.0488$. The corresponding overall PWER then equals 0.0434. Note that here the critical values are specifically chosen to produce the example, and not to control one of the error rates.

\section{Real data application} \label{sec:real_data}

We apply the PWER-P and PWER-U to a real data example, and compare the resulting critical values to the  PWER-adjusted, FWER-adjusted and unadjusted critical values. The data is taken from a data set created by \citet{kesselmeier2020} based on a randomized controlled trial which compares moxifloxacin and meropenem with meropenem alone in patients with severe sepsis or septic shock \cite{brunkhorst2012}. The primary endpoint is the mean daily total Sequential Organ Failure Assessment (SOFA) score over 14 days, which ranges from 0-24, with higher scores indicating worse outcome. \citet{kesselmeier2020} define two overlapping patient populations, 
\begin{align*}
\Pop_1 &= \text{patients with baseline lactate value $> 2$ mmol/L}\\
\Pop_2 &= \text{patients with baseline C-reactive protein value $> 128$ mg/L}
\end{align*}
and generate $1,000$ bootstrap samples from the trial data, for total patient numbers of $N = 100$, $200$ and $500$, respectively. 

For each bootstrap sample and error rate, we compute a common critical value for two right-sided tests performed in $\Pop_1$ and $\Pop_2$, assuming normally distributed outcomes with known and homogeneous variances (see Section~3.2 in \cite{luschei2025prevalence}). We further assume distinct treatments ($T_1 \neq T_2$) and a stratified randomization to the treatments. The sample sizes of the three strata, $\Pop_{\{1\}}$, $\Pop_{\{2\}}$ and $\Pop_{\{1,2\}}$ are used to derive the correlation between the test statistics and to estimate the prevalences, as will be described in Section~\ref{sec:pi-est}. Because the tests are one-sided, we use a significance level of $\alpha = 0.025$. Under the unadjusted testing, the critical value is $\Phi^{-1}(1 - \alpha) = 1.96$, where $\Phi^{-1}$ denotes the quantile function of the standard normal distribution. 

\begin{figure}[h]
  \centering
  \includegraphics[width=0.85\textwidth]{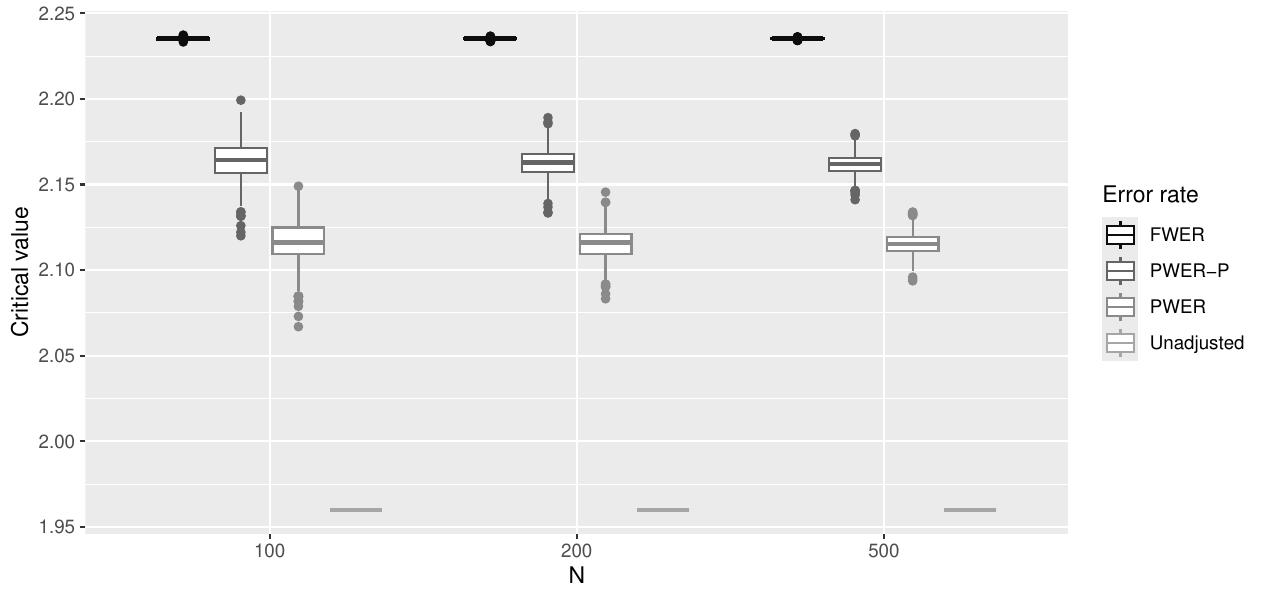}
  \caption{Critical values under control of the FWER, PWER and PWER-P in the real data example from \citet{kesselmeier2020} for different sample sizes $N$, with significance level $\alpha = 0.025$. In this example, the PWER-P and the PWER-U coincide since there are only two target populations.}
  \label{fig:real-ex}
\end{figure}

Figure~\ref{fig:real-ex} shows the resulting distributions of the critical values. As expected, all multiplicity-adjusted procedures have larger critical values than the unadjusted procedure and are thereby more conservative. The critical values obtained under PWER-P-control lie between those of the FWER- and PWER-adjusted procedures, being somewhat closer to the latter. Note that the PWER-P and PWER-U coincide in this setting, as only two target populations are considered. Consequently, their critical values are always identical.

\section{Power comparison} \label{sec:power}

In this section, we investigate the impact of applying the PWER-P or PWER-U on power and sample size. Therefore, we will extend the analyses presented in Section~5.2 of \citet{brannath2023}, which compared the performance of the PWER to the FWER. As in that setting, we consider a study with two overlapping populations $\Pop_i$, $i=1,2$, where an experimental treatment $T_i$ is compared to a common control treatment. Two possible scenarios are considered, (i) $T_1 \neq T_2$ and (ii) $T_1 = T_2$. To compare the power when using a common PWER-adjusted critical value $c_P$ versus an FWER-adjusted critical value $c_F$, \citet{brannath2023} introduced the quantity \[q_{\alpha, \beta}(c) = \left( \frac{\Phi^{-1}(1-\beta)+c}{\Phi^{-1}(1-\beta)+\Phi^{-1}(1-\alpha)}\right)^2, \quad \text{for } c \in \{c_P, c_F\},\] which represents the factor of sample size increase relative to unadjusted testing, when aiming for a marginal power of $1-\beta$ in each population. 

\begin{figure}[h]
\vspace{-0.75cm}
\centering
  \begin{minipage}[b]{0.47\textwidth}
    \makebox{\includegraphics[width=\textwidth]{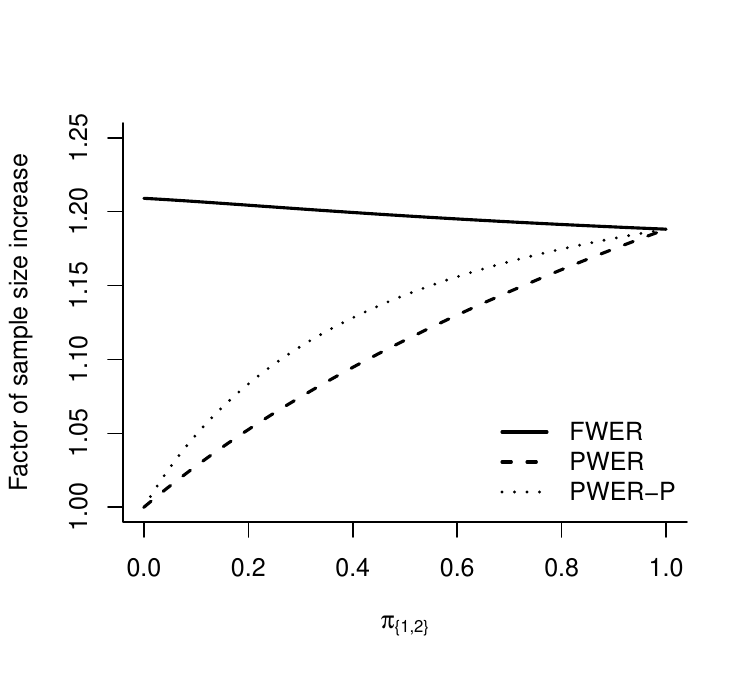}}
  \end{minipage}
  \begin{minipage}[b]{0.47\textwidth}
    \makebox{\includegraphics[width=\textwidth]{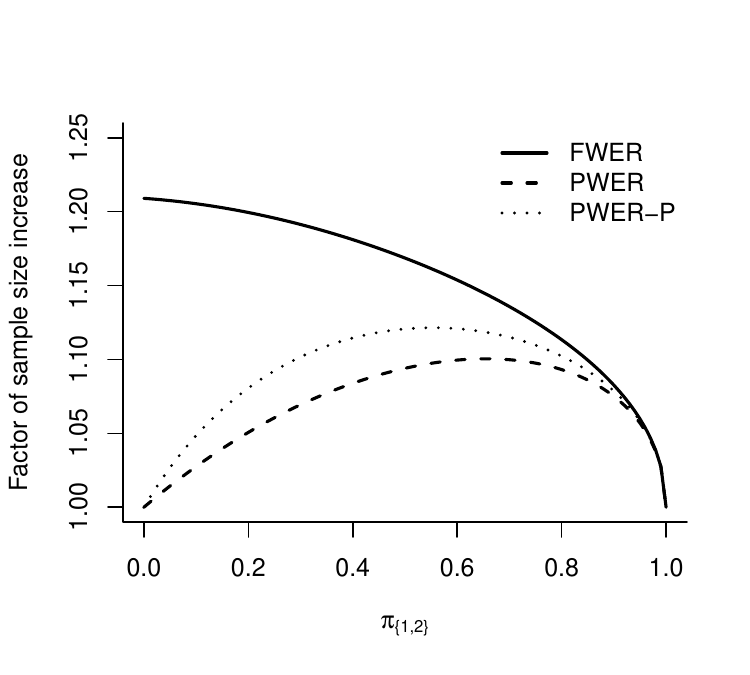}}
  \end{minipage}
\caption{\label{fig:power2pops} Factor of sample size increase compared to unadjusted testing for FWER-, PWER- and PWER-P-control at level $\alpha = 0.025$ in a study with two overlapping populations, depending on the intersection size $\pi_{\{1,2\}}$. The marginal power is $1-\beta = 0.8$. The left panel is for scenario (i) with different experimental treatments and a common control, the right panel is for scenario (ii) with equal experimental treatments.}
\end{figure}

Under the assumption that the strata $\Pop_{\{1\}}$ and $\Pop_{\{2\}}$ have equal size, \citet{brannath2023} plotted the two sample size increase factors $q_{\alpha, \beta}(c_P)$, $q_{\alpha, \beta}(c_F)$ as functions of the overlap prevalence $\pi_{\{1,2\}}$ for the scenarios (i) and (ii), assuming that $\alpha = 0.025$ and $\beta = 0.2$. They found that the PWER leads to smaller sample sizes than the FWER, with the difference decreasing as $\pi_{\{1,2\}}$ increases, until the values coincide at $\pi_{\{1,2\}} = 1$. Figure~\ref{fig:power2pops} shows these two plots, into which we have now added the corresponding curve for the PWER-P (which equals the PWER-U, since there are only two populations). It can be seen that, depending on $\pi_{\{1,2\}}$, the PWER-P leads to a slight sample size increase compared to the PWER, while still achieving a clear gain over the FWER.

Moreover, we extend these analyses to settings with three overlapping populations, where the PWER-P and PWER-U may differ. We therefore assume that the populations $\Pop_i$, $i=1,2,3$, are defined by independent binary biomarkers $B_i$, where $\Pop_i$ includes all patients expressing biomarker $B_i$. In a simulation loop, we randomly generate expression probabilities for the three biomarkers and use them to determine the prevalences of the strata $\Pop_J$, $J \subseteq \{1,2,3\}$, $J \neq \emptyset$. Then, we compute a common critical value $c$ for each of the four different error rates under the two scenarios (i) $T_i \neq T_j$, $i\neq j$ and (ii) $T_1 = T_2 = T_3$. Based on this $c$, we compute the sample size increase factors $q_{\alpha, \beta}(c)$ for $\alpha = 0.025$ and $\beta = 0.2$. This procedure is repeated $10,000$ times. The resulting distributions of sample size increase factors are shown in Figure~\ref{fig:power3pops}. Detailed summary statistics can be found in Table~\ref{tab:power} in Appendix~\ref{app2}. As in the setting with two populations, the PWER-P and PWER-U provide a good compromise between the PWER and the FWER. 

In scenario (i), the sample size increase factors obtained under the PWER-P and PWER-U are identical in all simulated studies. In scenario (ii), they differ in only 465 simulations (approximately 5\%), with all differences being negligible ($< 10^{-9}$), probably due to numerical imprecisions. Thus, these simulations reveal no meaningful differences between using the PWER-P and PWER-U.

We also conducted these analyses for $m=4$ and $m=5$ populations and obtained similar results. 

\begin{figure}[h]
\vspace{-0.75cm}
\centering
  \begin{minipage}[b]{0.47\textwidth}
    \makebox{\includegraphics[width=\textwidth]{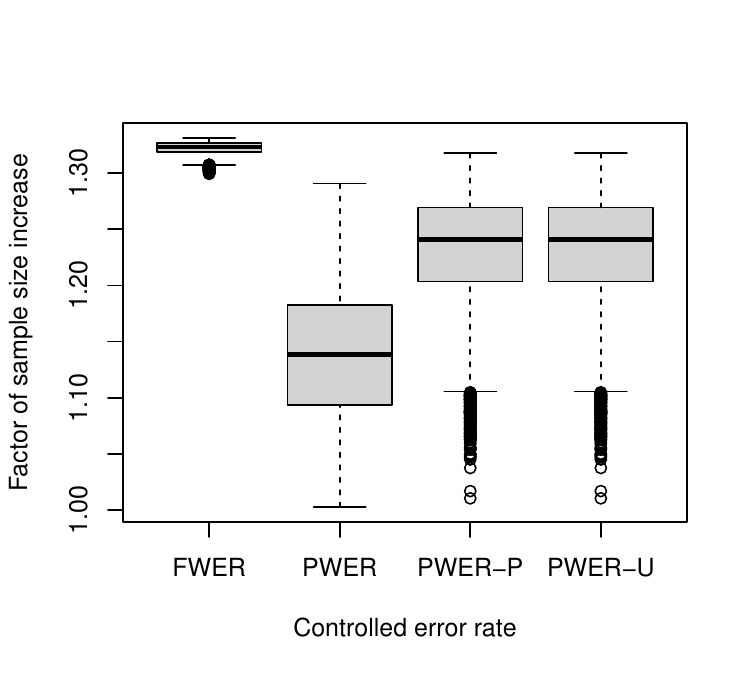}}
  \end{minipage}
  \begin{minipage}[b]{0.47\textwidth}
    \makebox{\includegraphics[width=\textwidth]{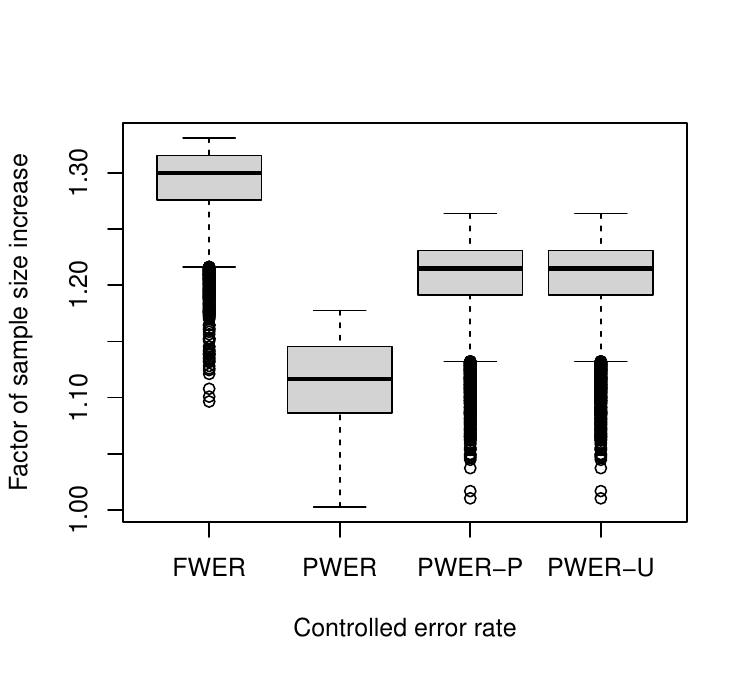}}
  \end{minipage}
\caption{\label{fig:power3pops} Factor of sample size increase compared to unadjusted testing for FWER-, PWER-, PWER-P- and \mbox{PWER-U}-control at level $\alpha = 0.025$ over $10,000$ different studies with three overlapping populations. The marginal power is $1-\beta = 0.8$. The left panel is for scenario (i) with $T_i \neq T_j$, $i \neq j$, the right panel is for scenario (ii) with $T_1 = T_2 = T_3$.}
\end{figure}

\section{Behavior of the maximal strata-wise FWER} \label{sec:maxfwer} 

An important question when controlling error rates such as the PWER, PWER-P and PWER-U is how large the maximal strata-wise FWER, $\max_{J \subseteq I: \, \pi_J \neq 0} \FWER_J$, can become. This value can be interpreted as the maximal probability for future patients to receive an inefficient treatment. For the PWER, this question has already been examined in \cite{brannath2023, luschei2025prevalence}. In different settings with many varying parameters (such as the number of populations, the distribution of the test statistics, etc), it was shown by simulations that controlling the PWER at level $\alpha$ often implies control of the maximal strata-wise FWER at a higher level, such as $2\alpha$. We now extend these simulations to the PWER-P and PWER-U. 

We again define $m$ populations $\Pop_1, \dots, \Pop_m$ based on binary biomarkers, where biomarker $i$ is expressed with a probability $p_i$, and $\Pop_i$ includes all patients who express the $i$-th biomarker. In every simulation run, the probabilities $p_i$ are sampled from independent uniform distributions and used to compute the prevalences of the strata via \[\pi_J = \frac{\prod_{j \in J} p_j \prod_{k \not\in J} (1-p_k)}{1-\prod_{i\in I} (1-p_i)}, \, J \subseteq I.\] We then determine the correlation matrix of the test statistics, for example using formula~(\ref{eq:corr}) in the single treatment case ($T_i = T_j$ for all $i,j \in I$). Subsequently, the prevalences and correlations are  used to compute critical values for PWER-, PWER-P- and PWER-U-control, as described in Sections~\ref{sec:intro} and \ref{sec:control}.

\begin{figure}[h]
  \centering
  \includegraphics[width=0.8\textwidth]{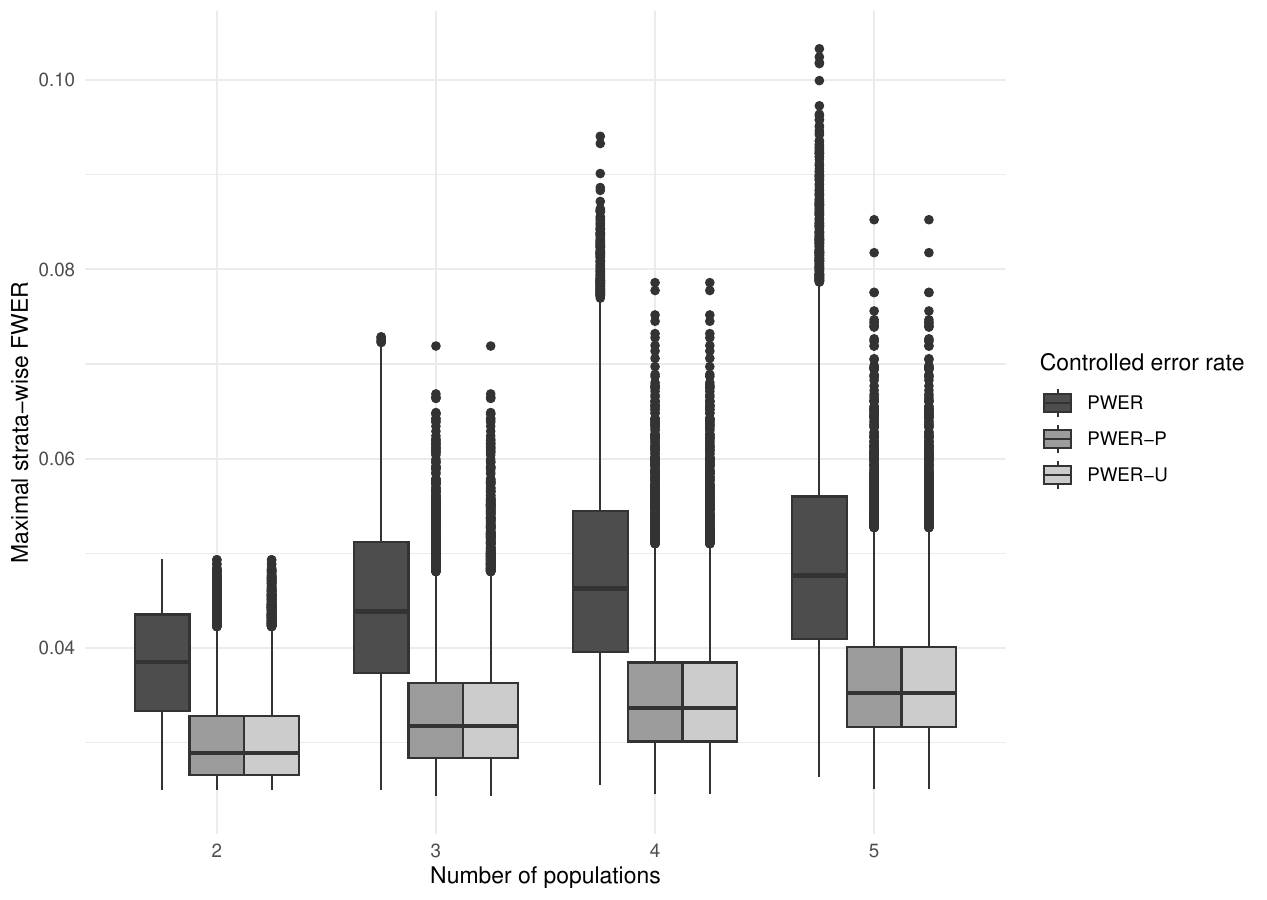}
  \caption{Distribution of the maximal strata-wise FWER under control of the PWER, PWER-P and PWER-U at level $\alpha = 0.025$ for $m\in \{2, \dots, 5\}$ populations.}
  \label{fig:max-fwer}
\end{figure}

Figure~\ref{fig:max-fwer} displays the resulting maximal strata-wise FWERs under the assumption of normally distributed test statistics and equal treatments. We considered $m \in \{2, \dots, 5\}$ populations and set the significance level to $\alpha = 0.025$. Each boxplot is based on $10,000$ simulation runs using identical random seeds. The results show that PWER-P- and PWER-U-control provide a clear improvement over the classical PWER-control. While the maximal strata-wise FWER under PWER control lies, on average, between 0.04 and 0.05, it is reduced to values between 0.03 and 0.037 under the PWER-P and PWER-U. The detailed summary statistics of the simulated values can be found in Table~\ref{tab:max-fwer-summary} in Appendix~\ref{app2}. We obtained similar results in the multiple treatments case ($T_i \neq T_j$ for $i \neq j$).

As in Section~\ref{sec:power}, the PWER-P and PWER-U showed almost no differences in these simulations. Deviations in the resulting maximal strata-wise FWERs only occurred in rare cases and were so small, that they were probably caused by  numerical imprecisions. Moreover, there was no instance where the PWER caused a smaller maximal strata-wise FWER than the PWER-P, although the condition~(\ref{cond1}) from Theorem~\ref{thm1} was sometimes violated (e.g.\ in 1701 simulation runs with equal treatments, $m=4$ and under PWER-control). But note that Theorem~\ref{thm1} is only applicable for fixed critical values, whereas we here calculate separate critical values to control each error rate.

In the simulations presented so far, all strata are assigned a prevalence greater than 0, even if it can be very close to 0. Therefore, we also extended the simulations to settings in which certain strata are entirely absent. In particular, we regarded settings where certain populations are disjoint from all other populations, such that the PWER-P-level may be utilized more efficiently by computing heterogeneous critical values (see Section~\ref{sec:disj_pop}). In these cases, the maximal strata-wise FWERs often become smaller (under all error rates), because strata belonging to many populations may disappear, and these strata generally have higher strata-wise FWERs than strata belonging to only few populations. Furthermore, the use of the PWER-P or PWER-U often results in comparatively smaller improvements of the maximal strata-wise FWER than that observed in the more general simulations described above. 

In general, since it is not possible to cover every single possible scenario in this work, we recommend that, in practice, for a concretely given study design, the maximal strata-wise FWER should always be investigated, for example through simulations.

\section{Estimation of strata prevalences} \label{sec:pi-est}

In practice, the prevalences $\pi_J$ of the strata $\Pop_J$ are usually unknown and therefore need to be estimated from the data. This raises the question whether suitable control of the PWER-P and PWER-U can then still be reached. In \cite{brannath2023, luschei2025prevalence}, we have already examined this problem for the PWER in detail, in the situation where the prevalences are estimated by their maximum-likelihood estimate (MLE) under a multinomial distribution. The MLE is given by 
\begin{align} \label{eq:pi-est}
\hat{\pi}_J = n_J/N, \quad J \subseteq I,
\end{align}
where $n_J$ denotes the observed sample size in the stratum $\Pop_J$, and $N$ denotes the total sample size of the study. Both asymptotic and finite-sample results showed that the true PWER remains well controlled when the critical values are computed from the estimated prevalences. 

The asymptotic result extends directly to the PWER-P and PWER-U. Since the MLE converges almost surely to the true prevalences, the true PWER-P and PWER-U converge almost surely towards the significance level $\alpha$ that is used for the control of the estimated PWER-P or PWER-U.

To investigate the finite-sample behaviour, we extend the simulation study of \cite{luschei2025prevalence} to the PWER-P and PWER-U. The simulation setup is identical to that described in Section~\ref{sec:maxfwer}. In each simulation run, we randomly generate a trial with overlapping target populations defined by binary biomarkers. For a given sample size $N$, we then generate the strata-wise sample sizes from the multinomial distribution $M(N, (\pi_J)_{J \subseteq I})$ and estimate the strata prevalences using Formula~(\ref{eq:pi-est}). The estimated prevalences are then used to compute the critical values, and the resulting true PWER, PWER-P and PWER-U are computed using the underlying true prevalences.

\begin{figure}[h]
  \centering
  \includegraphics[width=0.8\textwidth]{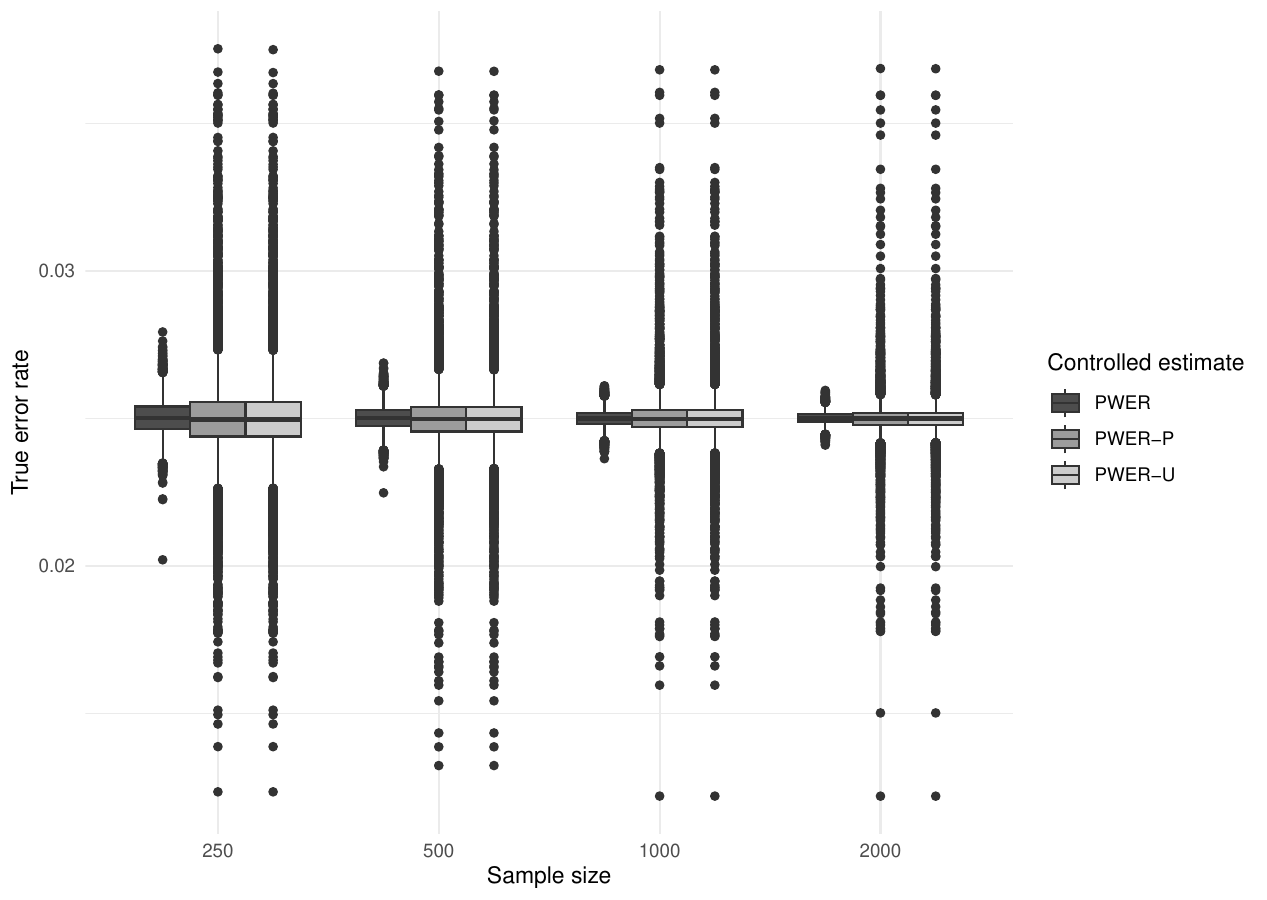}
  \caption{Distribution of the true PWER, PWER-P and PWER-U under control of the corresponding plug-in estimates in studies with $m=3$ populations, sample sizes $N \in \{250, 500, 1000, 2000\}$, and $\alpha = 0.025$.}
  \label{fig:true-pwer}
\end{figure}

Figure~\ref{fig:true-pwer} shows the distribution of the simulated true PWER, PWER-P and PWER-U for $10,000$ simulation runs with $m=3$ populations and different sample sizes $N \in \{250, 500, 1000, 2000\}$. For all three error rates, the mean and median closely match the targeted significance level of $\alpha=0.025$, showing that in the average, the plug-in estimates provides accurate control of the true error rates. Compared with the PWER, however, the distributions of the true PWER-P and PWER-U have a slightly larger variance. The summary statistics of the simulated values can be found in Table~\ref{tab:pi-est}.

For an individual study, it may therefore be of interest to quantify the uncertainty in the true PWER, PWER-P or PWER-U arising from the estimation of the strata prevalences. In \cite{luschei2026prediction}, an asymptotic prediction interval for the true PWER was derived using the delta method. It can easily be extended to the PWER-P and PWER-U by replacing the PWER with the corresponding error criterion.

\section{Extension to simultaneous confidence intervals} \label{sec:confint}

Similarly to the PWER, simultaneous confidence intervals can be derived for the PWER-P and PWER-U due to the correspondence between multiple testing and multi-dimensional confidence intervals (see e.g.\ \cite{finner1994, dickhaus2014}). We consider again one-sided null hypotheses of the form $H_i\colon \theta_i \leq 0$, where $\theta_i = \theta(T_i, \Pop_i)$ denotes the treatment effect of $T_i$ in population $\Pop_i$, $i=1, \dots, m$. As shown by \citet{brannath2023}, when using Wald-type test statistics of the form $\hat{\theta}_i/\text{SE}_i$ for $H_i$, where $\hat{\theta}_i$ is an estimate of $\theta_i$ with standard error $\text{SE}_i$, the corresponding simultaneous confidence intervals can be obtained by $\mathcal{C}_i = [\hat{\theta}_i -c_i \text{SE}_i, \infty)$. Here the critical values $c_i$ are those obtained from Equation (\ref{eq:pwerp-crit}) or (\ref{eq:pweru-crit}). Similarly, left-sided confidence intervals can be constructed by $\mathcal{C}_i = (-\infty, \hat{\theta}_i + c_i \text{SE}_i]$ for appropriately adjusted critical values. The resulting intervals then satisfy the coverage property
\[ \min_{i\in I^+} \sum_{J\in C_i} \frac{\pi_J}{\pi^{C_i}} \P\left(\theta_j\in \mathcal{C}_j \text{ for all } j \in J \right) \geq 1-\alpha \]
under PWER-P control and
\[ \min_{U \in \mathcal{U}^+} \sum_{J\in C_U} \frac{\pi_J}{\pi^{C_U}} \P\left(\theta_j\in \mathcal{C}_j \text{ for all } j \in J \right) \geq 1-\alpha \]
under PWER-U control. Thus, under PWER-P control, the average simultaneous coverage probability within each target population is at least $1-\alpha$, whereas under PWER-U control the same guarantee holds for every union of target populations.

\section{Summary and discussion}

We introduced two new multiple type I error rates for clinical trials with multiple target populations, the PWER-P and the PWER-U. While the original PWER can be interpreted as the probability that a randomly chosen future patient will receive an inefficient therapy, the idea of the PWER-P is to control this probability within all target populations. Thereby, unbalanced type I error allocations across the different populations can be prevented. The PWER-U extends this approach by considering all possible unions of target populations. This can be of interest e.g.\ in adaptive enrichment designs where the  populations for potential treatment approval are not determined from the beginning. One might then be interested in performing error control for all possible combinations of subpopulations. However, in the simulations conducted for this paper (for single stage designs) we found that the PWER-P- and PWER-U-test adjustments often coincide or are very close to each other. Extending the PWER-P and PWER-U to adaptive or group sequential designs is an important direction for future research.

The main advantage of the PWER-P or PWER-U over the PWER is that the use of a local (population-wise) error level of at most $\alpha$ is mandatory and part of the error criterion, whereas with the PWER it is only an option. Indeed, unbalanced error levels may be undesirable in many situations. For example, consider a basket trial enrolling patients with different cancer types that share a common molecular alteration. It is difficult to justify why patients with one cancer type should be penalized by being assigned a local error level greater than $\alpha$, simply because a more stringent error level is chosen for patients with another cancer type.

The examples and simulations presented in this paper show that the PWER-P and PWER-U often provide a good compromise between the often overly conservative FWER and the sometimes overly liberal PWER, both with respect to the achieved type I error control and power. However, in Section \ref{sec:nested-pops}, we have also seen that in trials with nested populations, these error rates coincide with the FWER, such that the overall PWER may be a more appropriate choice.

As discussed by \citet{brannath2023fwel}, the overall PWER can be represented as a family-wise expected loss (FWEL) in the framework introduced by \citet{maurer2023}. In contrast, the PWER-P and PWER-U are defined as maxima over several population-specific or union-specific error rates and therefore cannot be viewed as a single expected loss. They may nevertheless be interpreted as a minimax approach which controls the largest patient risk across all populations or population unions. Thus, although existing optimality results regarding FWEL control cannot be transferred directly, related formulations may still be possible to derive optimal decision rules under PWER-P or PWER-U control.

An alternative to PWER-P- or PWER-U-control could be to control the maximal strata-wise FWER that we investigated in Section~\ref{sec:maxfwer}. However, that may possibly coincide with FWER control, namely when the stratum $\Pop_I = \cap_{i=1}^m \Pop_i$ has a positive prevalence. Furthermore, due to the fact that the strata prevalences are completely ignored, this approach could in turn be overly conservative. Conducting seperate tests in all strata $\Pop_J$, $J \subseteq I$ with $\pi_J >0$ is also not recommended and often impossible, as the number of strata increases rapidly with the number of populations. Consequently, the available sample size per stratum may be too low to reach the desired power.

The simulations conducted for this paper were based on the assumption that the test statistics follow a multivariate normal distribution. This is often justified, for example under the different observation models presented in \cite{luschei2025prevalence}. In other cases, it is also often justified by the multivariate central limit theorem \cite{vandervaart1998}. However, these models rely on the assumption of homogeneous null effects, i.e.\ that the null hypothesis applies to all populations and their intersections. Otherwise, the distribution of the test statistics needs to be approximated, e.g.\ via a bootstrap procedure, as we have shown in \cite{luschei2025fwer}.

We finally note that in this paper we have only considered one-sided tests and confidence intervals. Although it would theoretically be possible to reach error control also for two-sided tests and to construct symmetric confidence intervals, this would lead to a direction error problem: Type-I error probabilities would also be controlled for situations in which an experimental treatment is falsely concluded to be inferior to the control treatment. However, this additional protection has limited practical relevance, at least from the regulator's perspective, since such findings would not result in regulatory approval of the treatment. Thus, one would generally overcorrect for multiple type I errors.

\section*{Acknowledgements}

The authors thank Dr.\ Miriam Kesselmeier for kindly providing the data for the real data application.

\bibliographystyle{unsrtnat-max3}
\bibliography{references}

@article{brannath2023,
  author  = {Brannath, W. and Hillner, C. and Rohmeyer, K.},
  title   = {The population-wise error rate for clinical trials with overlapping populations},
  journal = {Statistical Methods in Medical Research},
  year    = {2023},
  volume  = {32},
  number  = {2},
  pages   = {334--352}
}

@article{luschei2025prevalence,
  author  = {Luschei, R. and Brannath, W.},
  title   = {The effect of estimating prevalences on the population-wise error rate},
  journal = {Statistical Methods in Medical Research},
  year    = {2025},
  volume  = {34},
  number  = {2},
  pages   = {390--404}
}

@article{luschei2025fwer,
  author  = {Luschei, R. and Brannath, W.},
  title   = {Family-wise error rate control in clinical trials with overlapping populations},
  year    = {2025},
  note    = {arXiv:2511.09449}
}

@article{luschei2026prediction,
  author  = {Luschei, R. and Brannath, W.},
  title   = {A prediction interval for the population-wise error rate},
  year    = {2026},
  note    = {arXiv:2602.06828}
}

@article{kesselmeier2020,
  author    = {Kesselmeier, M. and Benda, N. and Scherag, A.},
  title     = {Effect size estimates from umbrella designs: Handling patients with a positive test result for multiple biomarkers using random or pragmatic subtrial allocation},
  journal   = {PLoS ONE},
  year      = {2020},
  volume    = {15},
  number    = {8},
  pages     = {1--24}
}

@article{brunkhorst2012,
  author    = {Brunkhorst, F. and Oppert, M. and Marx, G. and Bloos, F. and Ludewig, K. and Putensen, C. and Nierhaus, A. and Jaschinski, U. and Meier-Hellmann, A. and Weyland, A. and Gründling, M. and Moerer, O. and Riessen, R. and Seibel, A. and Ragaller, M. and Büchler, M. and John, S. and Bach, F. and Spies, C. and Reill, L. and Fritz, H. and Kiehntopf, M. and Kuhnt, E. and Bogatsch, H. and Engel, C. and Loeffler, M. and Kollef, M. and Reinhart, K. and Welte, T. and German Study Group Competence Network Sepsis (SepNet)},
  title     = {Effect of empirical treatment with moxifloxacin and meropenem vs meropenem on sepsis-related organ dysfunction in patients with severe sepsis: a randomized trial},
  journal   = {JAMA},
  year      = {2012},
  volume    = {307},
  number    = {22},
  pages     = {2390--2399}
}

@article{park2020,
  author    = {Park JJH and Hsu G and Siden EG and Thorlund K and Mills EJ},
  title     = {An overview of precision oncology basket and umbrella trials for clinicians},
  journal   = {CA Cancer J Clin},
  year      = {2020},
  volume    = {70},
  number    = {2},
  pages     = {125--137}
}

@article{stallard2019,
  author    = {Stallard N and Todd S and Parashar D and Kimani PK and Renfro LA},
  title     = {On the need to adjust for multiplicity in confirmatory clinical trials with master protocols},
  journal   = {Ann Oncol},
  year      = {2019},
  volume    = {30},
  number    = {4},
  pages     = {506--509}
}

@article{baldi2023,
  author  = {Baldi Antognini, A. and Frieri, R. and Zagoraiou, M.},
  title   = {New insights into adaptive enrichment designs},
  journal = {Statistical Papers},
  year    = {2023},
  volume  = {64},
  pages   = {1305--1328}
}

@book{dickhaus2014,
  author    = {Dickhaus, T.},
  title     = {Simultaneous Statistical Inference: With Applications in the Life Sciences},
  publisher = {Springer},
  address   = {Berlin, Heidelberg},
  year      = {2014}
}

@article{finner1994,
  author  = {Finner, H.},
  title   = {Testing Multiple Hypotheses: General Theory, Specific Problems, and Relationships to Other Multiple Decision Procedures},
  note    = {Universität Trier, Habilitationsschrift},
  year    = {1994}
}

@book{vandervaart1998,
  author    = {Vaart, A. W. van der},
  title     = {Asymptotic Statistics},
  series    = {Cambridge Series in Statistical and Probabilistic Mathematics},
  publisher = {Cambridge University Press},
  address   = {Cambridge},
  year      = {1998}
}

@article{maurer2023,
  author  = {Maurer, W. and Bretz, F. and Xun, X.},
  title   = {Optimal test procedures for multiple hypotheses controlling the familywise expected loss},
  journal = {Biometrics},
  year    = {2023},
  volume  = {79},
  number  = {4},
  pages   = {2781--2793},
  doi     = {10.1111/biom.13907}
}

@article{brannath2023fwel,
  author  = {Brannath, W.},
  title   = {Discussion on ``Optimal test procedures for multiple hypotheses controlling the familywise expected loss'' by Willi Maurer, Frank Bretz, and Xiaolei Xun},
  journal = {Biometrics},
  year    = {2023},
  volume  = {79},
  number  = {4},
  pages   = {2806--2810},
  doi     = {10.1111/biom.13909}
}

\appendix
\renewcommand{\thesubsection}{\Alph{subsection}}
\section*{Appendix} 
\addcontentsline{toc}{section}{Appendix}

\subsection{Proof of Theorem \ref{thm1}} \label{app1}

The inequalities $\PWERP \leq \PWERU \leq \FWER$ follow directly from the respective definitions and the fact that $\FWER_J \leq \FWER$ for every $J \subseteq I$. It remains to prove that $\PWER \leq \PWERP$ holds under condition (\ref{cond1}). Define $p_J = \pi_J\FWER_J/\PWER$ and $p_i = \sum_{J \in C_i} p_J$. Since $\PWERP = \max_{i \in I^+} p_i\PWER/\pi^{C_i}$, it suffices to show that there exists an $i\in I^+$ with $p_i \geq \pi^{C_i}$. Assume, for contradiction, that $p_i < \pi^{C_i}$ for all $i \in I^+$. Summing over $i$ gives
\begin{align} \label{eq:p-pi}
\sum_{i \in I^+} p_i < \sum_{i\in I^+}\pi^{C_i} \quad \Longleftrightarrow \quad\sum_{J \subseteq I} |J| p_J < \sum_{J \subseteq I} |J|\pi_J.
\end{align}
For each $i \in I$, let $\pi_{|J| \geq i} = \sum_{J \subseteq I: \,|J| \geq i} \pi_J$. If $\pi_{|J| \geq i} > 0$, we define $\PWER_{|J| \geq i} = \sum_{J \subseteq I: \,|J| \geq i} \pi_J \FWER_J/\pi_{|J| \geq i}$, and otherwise $\PWER_{|J| \geq i} = 0$. Analogously, we also define $\pi_{|J| <i}$ and $\PWER_{|J|<i}$ with the condition $|J|\geq i$ replaced by $|J|<i$. By condition~(\ref{cond1}), 
\begin{align*}
\PWER_{|J| < i} \leq \max_{J \subseteq I: \, \pi_J \neq 0, \, |J| < i} \FWER_J \leq \min_{J \subseteq I: \, \pi_J \neq 0, \, |J| \geq i} \FWER_J \leq \PWER_{|J| \geq i}.
\end{align*}
Hence,
\[ \PWER = \pi_{|J|<i}\PWER_{|J| < i} + \pi_{|J| \geq i}\PWER_{|J| \geq i} \le \PWER_{|J| \geq i}, \]
where we use that $\pi_{|J| < i}+\pi_{|J| \geq i}=1$.
Finally, we obtain,
\begin{align*}
\sum_{J \subseteq I}|J|p_J = \sum_{J \subseteq I} \frac{|J|\pi_J\FWER_J}{\PWER} = \sum_{i \in I} \frac{\PWER_{|J| \geq i}\pi_{|J|\geq i}}{\PWER}  \geq \sum_{i \in I} \pi_{|J|\geq i} = \sum_{J \subseteq I} |J|\pi_J,
\end{align*}
which contradicts the claim made in (\ref{eq:p-pi}).

\subsection{Summary statistics for the simulated values} \label{app2}

\begin{table}[H]
\centering
\caption{Summary statistics for the sample size increase factors (Figure~\ref{fig:power3pops})}
\label{tab:power}
\begin{tabular}{ccccccccc}
\toprule
Scenario & Error rate & Mean & SD & Min & Q1 & Median & Q3 & Max \\
\midrule
\multirow{4}{*}{(i)} &FWER & 1.3220 & 0.0056 & 1.2991 & 1.3187 & 1.3231 & 1.3264 & 1.3312 \\
&PWER & 1.1390 & 0.0591 & 1.0026 & 1.0935 & 1.1382 & 1.1826 & 1.2904 \\
&PWER-P & 1.2318 & 0.0482 & 1.0105 & 1.2035 & 1.2406 & 1.2692 & 1.3175 \\
&PWER-U & 1.2318 & 0.0482 & 1.0105 & 1.2035 & 1.2406 & 1.2692 & 1.3175 \\
\midrule
\multirow{4}{*}{(ii)} &FWER  & 1.2910 & 0.0321 & 1.0966 & 1.2757 & 1.2998 & 1.3153 & 1.3311 \\
&PWER  & 1.1136 & 0.0390 & 1.0026 & 1.0864 & 1.1165 & 1.1456 & 1.1776 \\
&PWER-P & 1.2070 & 0.0350 & 1.0105 & 1.1915 & 1.2147 & 1.2310 & 1.2637 \\
&PWER-U & 1.2070 & 0.0350 & 1.0105 & 1.1915 & 1.2147 & 1.2310 & 1.2637 \\
\bottomrule
\end{tabular}
\end{table}

\begin{table}[H]
\centering
\caption{Summary statistics for the maximal strata-wise FWER (Figure~\ref{fig:max-fwer})}
\label{tab:max-fwer-summary}
\begin{tabular}{ccccccccc}
\toprule
$m$ & Error rate & Mean & SD & Min & Q1 & Median & Q3 & Max \\
\midrule
2 & PWER   & 0.0383 & 0.0063 & 0.0250 & 0.0333 & 0.0385 & 0.0435 & 0.0494 \\
2 & PWER-P & 0.0303 & 0.0048 & 0.0250 & 0.0265 & 0.0289 & 0.0328 & 0.0493 \\
2 & PWER-U & 0.0303 & 0.0048 & 0.0250 & 0.0265 & 0.0289 & 0.0328 & 0.0493 \\
\midrule
3 & PWER   & 0.0447 & 0.0096 & 0.0250 & 0.0373 & 0.0439 & 0.0512 & 0.0729 \\
3 & PWER-P & 0.0332 & 0.0064 & 0.0243 & 0.0284 & 0.0318 & 0.0363 & 0.0719 \\
3 & PWER-U & 0.0332 & 0.0064 & 0.0243 & 0.0284 & 0.0318 & 0.0363 & 0.0719 \\
\midrule
4 & PWER   & 0.0477 & 0.0110 & 0.0255 & 0.0396 & 0.0463 & 0.0545 & 0.0941 \\
4 & PWER-P & 0.0351 & 0.0068 & 0.0245 & 0.0301 & 0.0337 & 0.0385 & 0.0786 \\
4 & PWER-U & 0.0351 & 0.0062 & 0.0245 & 0.0301 & 0.0337 & 0.0385 & 0.0786 \\
\midrule
5 & PWER   & 0.0494 & 0.0116 & 0.0264 & 0.0409 & 0.0476 & 0.0560 & 0.1030 \\
5 & PWER-P & 0.0366 & 0.0070 & 0.0251 & 0.0317 & 0.0352 & 0.0401 & 0.0852 \\
5 & PWER-U & 0.0366 & 0.0070 & 0.0251 & 0.0317 & 0.0352 & 0.0401 & 0.0852 \\
\bottomrule
\end{tabular}
\end{table}

\begin{table}[H]
\centering
\caption{Summary statistics for the true PWER, PWER-P and PWER-U under control of the corresponding estimates (Figure~\ref{fig:true-pwer})}
\label{tab:pi-est}
\begin{tabular}{ccccccccc}
\toprule
$N$ & Error rate & Mean & SD & Min & Q1 & Median & Q3 & Max \\
\hline
250  & PWER   & 0.0250 & 0.0006 & 0.0202 & 0.0246 & 0.0250 & 0.0254 & 0.0279 \\
250  & PWER-P & 0.0250 & 0.0016 & 0.0123 & 0.0244 & 0.0250 & 0.0256 & 0.0375 \\
250  & PWER-U & 0.0250 & 0.0016 & 0.0123 & 0.0244 & 0.0250 & 0.0256 & 0.0375 \\
\midrule
500  & PWER   & 0.0250 & 0.0004 & 0.0225 & 0.0247 & 0.0250 & 0.0253 & 0.0269 \\
500  & PWER-P & 0.0250 & 0.0012 & 0.0132 & 0.0246 & 0.0250 & 0.0254 & 0.0368 \\
500  & PWER-U & 0.0250 & 0.0012 & 0.0132 & 0.0246 & 0.0250 & 0.0254 & 0.0368 \\
\midrule
1000 & PWER   & 0.0250 & 0.0003 & 0.0236 & 0.0248 & 0.0250 & 0.0252 & 0.0261 \\
1000 & PWER-P & 0.0250 & 0.0009 & 0.0122 & 0.0247 & 0.0250 & 0.0253 & 0.0368 \\
1000 & PWER-U & 0.0250 & 0.0009 & 0.0122 & 0.0247 & 0.0250 & 0.0253 & 0.0368 \\
\midrule
2000 & PWER   & 0.0250 & 0.0002 & 0.0241 & 0.0249 & 0.0250 & 0.0251 & 0.0260 \\
2000 & PWER-P & 0.0250 & 0.0007 & 0.0122 & 0.0248 & 0.0250 & 0.0252 & 0.0368 \\
2000 & PWER-U & 0.0250 & 0.0007 & 0.0122 & 0.0248 & 0.0250 & 0.0252 & 0.0368 \\
\bottomrule
\end{tabular}
\end{table}

\end{document}